\documentclass[twocolumn,preprintnumbers,amsmath,amssymb]{revtex4}

\usepackage{graphicx}
\usepackage{dcolumn}
\usepackage{bm}

\usepackage{color}
\usepackage{placeins}

\usepackage[normalem]{ulem}
\usepackage[colorlinks]{hyperref}

\begin{document}

\title{
Magnetoluminescence of ZnMnSe/BeMnTe heterostructures with type-II band alignment at millikelvin temperatures 
}

\author{ Dennis Kudlacik$^1$, Linda Kersting$^1$, Nataliia E. Kopteva$^1$, Mladen Kotur$^1$,  Dmitri R. Yakovlev$^{1}$,  and Manfred Bayer$^{1,2}$}
\affiliation{$^1$ Experimentelle Physik 2, Technische Universit\"at Dortmund, 44227 Dortmund, Germany}
\affiliation{$^2$ Research Center FEMS, Technische Universit\"at Dortmund, 44227 Dortmund, Germany}

\date{\today}

\begin{abstract}
The magneto-optical properties of a Zn$_{0.99}$Mn$_{0.01}$Se/Be$_{0.93}$Mn$_{0.07} $Te diluted magnetic semiconductor heterostructure with type-II band alignment are investigated at cryogenic temperatures down to 16 mK. The temperature of the Mn spin system, which at the lowest possible laser power reaches 270~mK,  is evaluated from the giant Zeeman splitting of the direct exciton in the Zn$_{0.99}$Mn$_{0.01}$Se layers subject to an external magnetic field. The degree of circular polarization of the direct and indirect optical transitions, induced by the magnetic field, is a sensitive indicator for the laser heating of the Mn spin system.  Evidence of spin glass formation in the Mn spin system of the Be$_{0.93}$Mn$_{0.07}$Te layers with the critical temperature of $T_{SG}=400$~mK is found.
\end{abstract}

\maketitle

\section{Introduction}

Diluted magnetic semiconductors  (DMS) based on II-VI semiconductors, like (Cd,Mn)Te, (Zn,Mn)Se, (Hg,Mn)Te, etc. are well established materials with bright magneto-optical properties~\cite{DMS_book_1988,DMS_book,Furdyna1988}. These properties are provided by the implementation of the  Mn$^{2+}$ magnetic ions, which localized magnetic moments strongly coupled to the conduction band electrons and valence band holes. This results in giant magneto-optical effects, like giant Zeeman splitting, giant Faraday/Kerr rotation, formation of magnetic polarons, etc. 

In turn, the magneto-optical effects give access to the investigation of spin interactions and spin dynamics in a system of localized Mn spins. Depending on the Mn concentration, which can be tuned across the full range from 0 to 100\%, and on the lattice temperature, the Mn spin system can be in a paramagnetic, spin-glass, or antiferromagnetic phase. This varying phenomenology is provided by the antiferromagnetic interactions between the neighboring Mn spins. For low Mn concentrations of about 1\%, small clusters, e.g., pairs of localized Mn spins, control the spin dynamics and in particular the spin-lattice relaxation (SLR) rate~\cite{Strutz1992,Debus2016}. For example, a strong dependence of the SLR time on the Mn concentration covering more than six orders of magnitude from tens of nanoseconds up to milliseconds has been found and related to the Mn-Mn interactions~\cite{Farah1996,Scalbert1996,Kneip2006,DMS_book_Chap_8}.  

It is instructive to study DMS with a low Mn concentration at very low temperatures, below the pumped liquid helium temperature of $T=1.6$~K, where the thermal energy is smaller than the Mn-Mn interaction energy, especially when the Mn-ions are not in nearest neighbor positions. Not many of such studies have been reported, especially not those involving magneto-optical techniques, as light illumination generates photocarriers, which can transfer their kinetic energy to the Mn spin system and therefore heat it~\cite{Keller2001,DMS_book_Chap_8}. Most reported experimental data on magnetic interactions in the Mn spin system have been obtained from measurements of the magnetic susceptibility via SQUID and of the specific heat. An overview can be found in Ref.~\onlinecite{Oseroff_Ch3_1988}. Information on the spin glass transition temperature in Mn-based II-VI DMS measured down to 10~mK can be found for  Cd$_{1-x}$Mn$_x$Te~\cite{Novak1985,Novak1986},  Cd$_{1-x}$Mn$_x$Se~\cite{Novak1984,Novak1986}, and Hg$_{1-x}$Mn$_x$Te~\cite{Brandt1983}, see also Ref.~\onlinecite{Rigaux_Ch6_1988} for an overview. We are aware of only one paper reporting photoluminescence of a two-dimensional electron gas system in a Cd$_{0.999}$Mn$_{0.001}$Te quantum well (QW) at $T=450$~mK~\cite{Imanaka2002}. A few further papers address the transport properties of Cd$_{1-x}$Mn$_x$Se layers~\cite{Glod1994} and Cd$_{1-x}$Mn$_x$Te QWs at millikelvin temperatures, including the observation of quantum Hall~\cite{Jaroszynski2000} and fractional quantum Hall~\cite{Betthausen2014} effects.

Heterostructures with type-II band alignment can be realized using the II–VI semiconductor combination ZnSe/BeTe, which serves as an excellent testbed system. In both materials, Mn$^{2+}$ magnetic ions can be incorporated at high concentrations, without degradation of their optical properties, facilitating the study of several giant magneto-optical effects typical of DMS in (Zn,Mn)Se/(Be,Mn)Te heterostructures. On this basis, we previously implemented the concept of heteromagnetic heterostructures with different Mn concentrations in the QWs and barrier layers, which allows one to accelerate the spin-lattice relaxation of Mn$^{2+}$ spins in layers with low Mn concentration by means of spin diffusion and faster relaxation in layers with high Mn concentration~\cite{Scherbakov2005}. This is a well suited approach to minimize the effect of laser heating on the Mn spin system. It motivates us to choose for the present study such a heteromagnetic structure with 1\% of Mn in the (Zn,Mn)Se layers (hosting the direct optical transition) and 7\% of Mn in the (Be,Mn)Te layers. 

In this paper, we study the magneto-optical properties of a Zn$_{0.99}$Mn$_{0.01}$Se/Be$_{0.93}$Mn$_{0.07} $Te DMS heterostructure with type-II band alignment at cryogenic temperatures down to 16~mK. The polarized photoluminescence of direct and indirect optical transitions in this system is measured in an external magnetic field, and information on the Mn spin system is extracted from these data. We analyze the experimental conditions required to minimize optical heating of the Mn spin system and to reach its lowest obtainable temperature of 270~mK in our study. 

The paper is organized as follows. Information on the studied sample and a description of the experimental setup are given in Sec.~\ref{sec:Experimentals}. Results of the magneto-optical studies of the direct optical transition are presented in  Sec.~\ref{sec:Direct}, while in Sec.~\ref{sec:Indirect} experimental data for the indirect optical transition are collected. 


\section{Experimentals}
\label{sec:Experimentals}

\subsection{Sample}

The sample used in this study is a DMS quantum well structure Zn$_{0.99}$Mn$_{0.01}$Se/Be$_{0.93}$Mn$_{0.07}$Te with type-II band alignment (sample code cb1748). It was grown by molecular-beam epitaxy on a (100)-oriented GaAs substrate. It contains 10 structural periods, each consisting of a 20-nm-thick Zn$_{0.99}$Mn$_{0.01}$Se layer and a 10-nm-thick Be$_{0.93}$Mn$_{0.07}$Te layer.

\begin{figure}[hbt]
	\begin{center}
\includegraphics[width=0.30\textwidth]{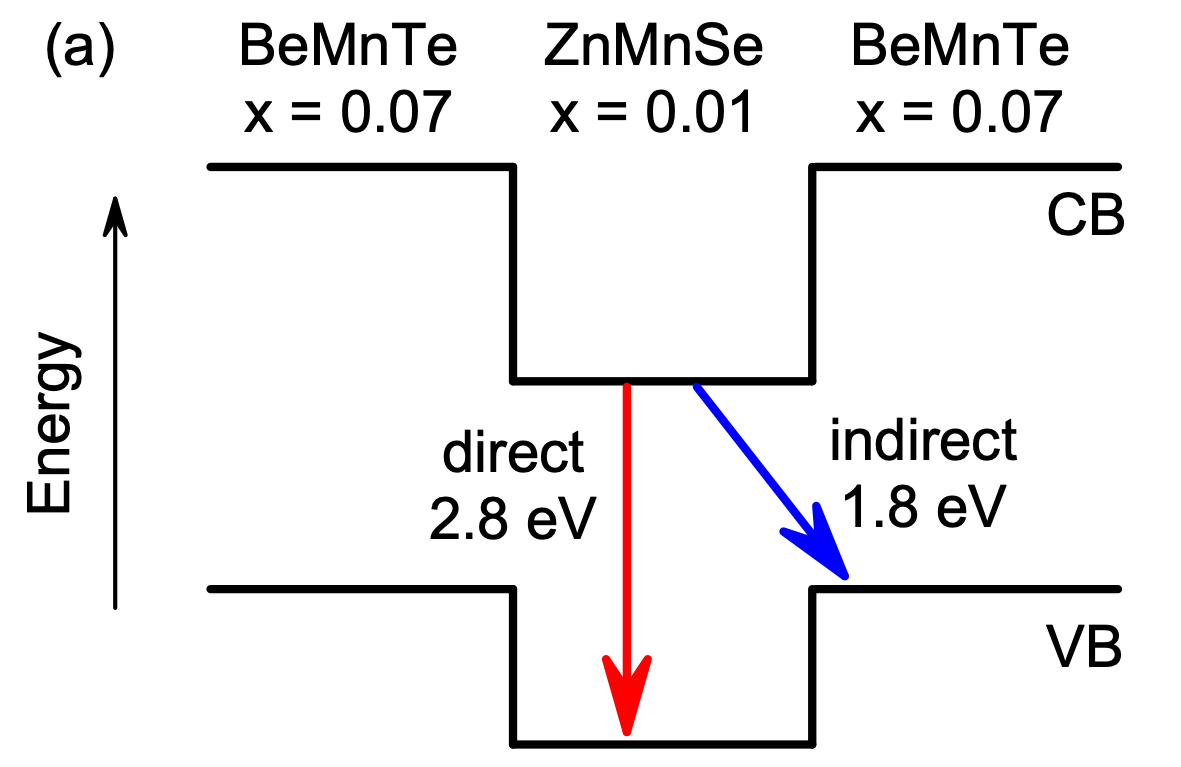}
\includegraphics[width=0.5\textwidth]{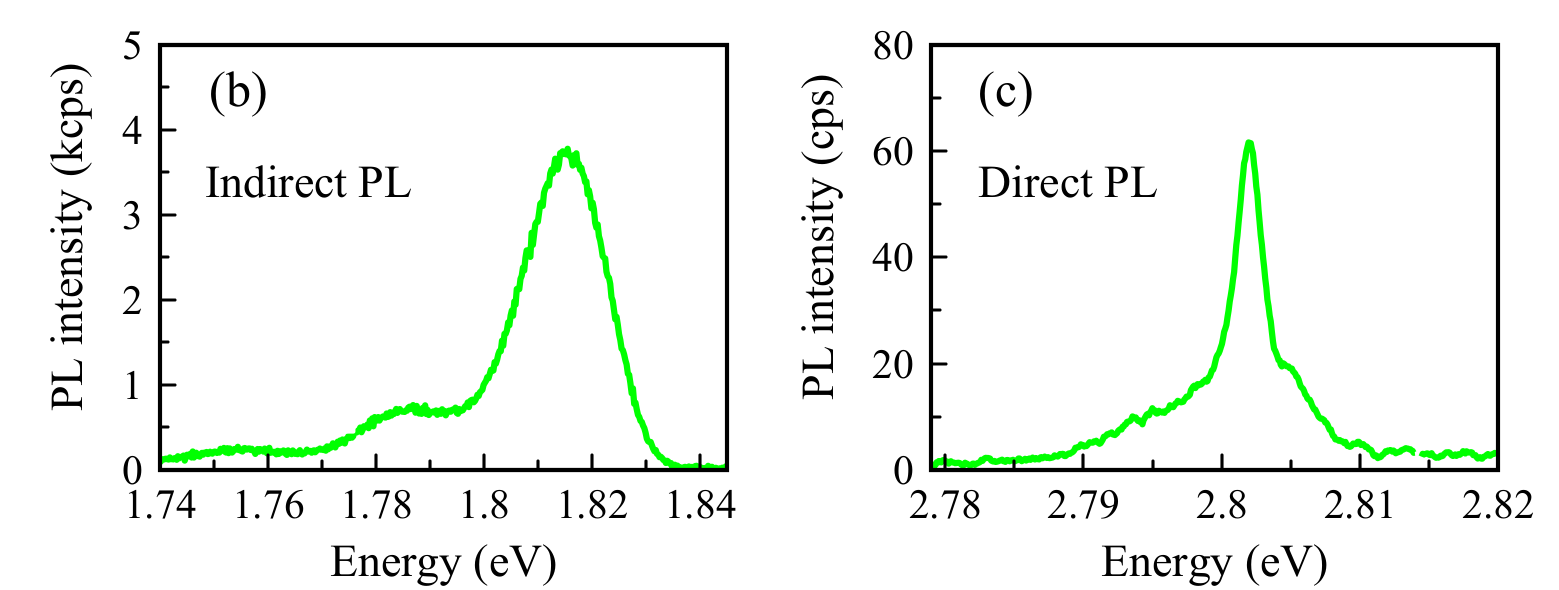}
\caption{(a) Band structure of the Zn$_{0.99}$Mn$_{0.01}$Se/Be$_{0.93}$Mn$_{0.07}$Te DMS quantum wells in the studied heterostructure. The band gap of (Zn,Mn)Se is 2.82~eV and that of (Be,Mn)Te is 4.2~eV. 
The direct (red arrow) and indirect (blue arrow) in real space optical transitions are shown.  (b,c) Photoluminescence  spectra of the indirect and direct transition emission measured at $T_s=16$~mK using $P=4$~$\mu$W excitation power.
}
\label{Fig_structure}
	\end{center}
\end{figure}

The ZnSe/BeTe heterostructures with type-II band alignment feature large band offsets in the conduction band of 2.2~eV and in the valence band of 0.8~eV, see Fig.~\ref{Fig_structure}(a). The conduction band electrons have their energy minimum in the ZnSe layer, while valence band holes have their minimum in the BeTe layer~\cite{Platonov1998}. The direct band gap of ZnSe is 2.82~eV and that of BeTe is 4.2~eV, resulting in an indirect band gap of about $1.8-2.0$~eV in the ZnSe/BeTe heterostructure. The optical and magneto-optical properties of ZnSe/BeTe heterostructures have been studied in detail in Refs.~\cite{Platonov1998,Platonov1999,Yakovlev2000,Maksimov2000,Maksimov2006}. Adding a small concentration of Mn results in a slight increase of the band gaps without having no significant effect on the band structure, and causing only minor shifts of the energies of the direct and indirect optical transitions and the related photoluminescence lines. There are a few manuscripts reporting on the magneto-optical properties of (Zn,Mn)Se/(Be,Mn)Te DMS QWs~\cite{Yakovlev2001,Yakovlev2002,Scherbakov2005,Kneip2006,Debus2010,Maksimov2019}.

Quantum wells with type-I band alignment containing (Zn,Mn)Se DMS layers, such as (Zn,Mn)Se/(Zn,Be)Se or (Zn,Mn)Se/(Zn,Be,Mg)Se, have been studied in more detail~\cite{Koenig1999,Keller2001,Akimov2006,Debus2016}. Comprehensive information on the energy transfer between the photogenerated charge carriers, the magnetic ion spin system and the lattice (phonon system) has been collected, for a review see Ref.~\cite{DMS_book_Chap_8}. Also, the problem of heating the Mn spin system by photogenerated carriers has been studied in detail.  

\subsection{Experimental setup}

The magneto-optical experiments at cryogenic temperatures down to 16~mK are performed with the use of a Proteox MX cryogenic system (Oxford Instruments). The system comprises a cryofree dilution refrigerator with a He4/He3 mixture. The cold finger temperature can be varied in the range from 16~mK up to 30~K. The sample was mounted on a copper puck thermally anchored to the mixing chamber plate of the cryostat, with the base temperature reached without sample illumination of $T_m=10\pm2$~mK. The cryostat is equipped with optical windows, allowing direct optical excitation and detection. The temperature was monitored using two calibrated sensors, one positioned on the mixing plate ($T_m$) and another one placed on the sample holder in close proximity to the sample ($T_s$). The sample holder is made of a cold finger fabricated from oxygen free copper (ultra-pure copper typically is featured by $>$99.99\% copper with a minimal oxygen content of $<$0.001\%) and has a thickness of 2~mm. A ruthenium oxide temperature sensor is attached to the back side of the cold finger, opposite the sample.   In order to avoid heating induced by the current sent through the temperature sensor, we use short measurements with a read out time of about 1~s, while applying the lowest possible current. In our experiments with no or weak laser excitation, the lowest temperature reached was $T_s=16$~mK.

The cryostat is equipped with a superconducting vector magnet, generating fields up to $B_{z}=5$~T in the Faraday geometry (field parallel to the light wave vector) and up to $B_{x}=B_{y}=1$~T in the Voigt geometry. In the reported experiments, the magnetic field is applied only in the Faraday geometry. 

For optical experiments, a lens with a focal length of 10~mm is placed on the sample holder at a distance of 10~mm from the sample. It is used for both laser excitation and collection of the sample emission. The diameter of the excitation laser spot on the sample is approximately 50~$\mu$m. The photoluminescence (PL) is excited using a continuous-wave diode laser with a photon energy of 3.06~eV (wavelength of 405~nm). The laser output is coupled into the setup using an optical fiber, after which we convert it to a parallel free beam. The laser light is unpolarized, and its power was varied from 2~nW up to 4~$~\mu$W to keep it at the lowest possible level, to minimize the effects of heating the sample lattice (i.e. the phonon system) and the Mn spin system. The PL is sent through polarization optics ($\lambda$/4 waveplate and Glan-Thompson linear polarizer) in order to analyze its circular polarization. For recording the PL spectra, a 0.5-m focal length Acton spectrometer (grating 1800~groves/mm) interfaced with a liquid-nitrogen-cooled silicon charge-coupled-device (CCD) detector is used. The typical signal accumulation times range from 1 hour (for the minimal excitation power of 2~nW) down to 10~s (for 4~$\mu$W).


\section{Experimental results}
\label{sec:Results}

In the studied heteromagnetic heterostructure, Zn$_{0.99}$Mn$_{0.01}$Se/Be$_{0.93}$Mn$_{0.07}$Te with type-II band alignment, the laser light with a photon energy of 3.06~eV is absorbed only in the (Zn,Mn)Se layers, as the band gap of (Be,Mn)Te exceeds the laser energy. After photogeneration, the electrons remain in (Zn,Mn)Se, while most of the holes scatter into (Be,Mn)Te, where they can reach minimum energy, see Fig.~\ref{Fig_structure}(a). These space-separated carriers contribute to the indirect PL with maximum at 1.815~eV, see Fig.~\ref{Fig_structure}(b). In the studied structure, the (Zn,Mn)Se layers have a width of 20~nm and the confined electrons create a Coulomb potential for the holes with their local energy minimum in the (Zn,Mn)Se layers~\cite{Platonov1998}. As a result, during short times after photogeneration of about 10~ps direct PL is generated~\cite{Maksimov2006}, which we measure here under continuous wave excitation and time-integrated detection at 2.802~eV, see Fig.~\ref{Fig_structure}(c). 

Diluted magnetic semiconductors with Mn$^{2+}$ magnetic ions are well known for their giant magneto-optical effects provided by the strong exchange interaction with the electronic band states subject to $sp-d$ hybridization~\cite{DMS_book_1988,DMS_book}. By means of polarized PL, one can measure the giant Zeeman splitting of the exciton states ($\Delta E_Z$) and the degree of circular polarization (DCP, $P_c$). The first effect gives information on the Mn spin system, namely on the Mn spin temperature $T_{Mn}$. The second effect is contributed by Mn spin temperature, but also by the spin dynamics of excitons or recombining charge carriers, which in DMS at cryogenic temperatures are contributed by Mn spin fluctuations and magnetic polaron formation~\cite{DMS_book_Chap_7}. Both of these effects are exploited in our experiments. 

The giant Zeeman splitting of the exciton states or of the band gap is proportional to the magnetization and thus to the average spin of the Mn ions $\langle S_z \rangle$
\begin{equation}
\Delta E_Z= (\alpha - \beta) N_0 x \langle S_z \rangle.
\label{eq:GZS1}
\end{equation}
Here, $N_0 \alpha$ and $N_0 \beta$ are the exchange constants for the conduction and valence band, respectively. 
In Zn$_{1-x}$Mn$_x$Se, $N_0 \alpha = 0.26$~eV and $N_0 \beta = -1.31$~eV~\cite{Twardowski1984}, while in Be$_{1-x}$Mn$_x$Te  $N_0 \beta = -0.4$~eV~\cite{Yakovlev2001}. $N_0$ is the inverse unit-cell volume and $x$ is the Mn mole fraction. $\langle S_z \rangle$ is the mean thermal value of the Mn spin component along the magnetic field $\textbf{B}$ at the Mn spin temperature $T_{Mn}$. It can be expressed by the Brillouin function ${\rm B}_{5/2}$:
\begin{equation}
\langle S_z \rangle = -S_{eff}(x) {\rm B}_{5/2} \left[ \frac{5 g_{Mn} \mu_B B}{2 k_B (T_{Mn} + T_0(x))} \right].
\label{eq:GZS2}
\end{equation}
Here, $g_{Mn}=2 $ is the $g$-factor of the Mn$^{2+}$ ions. $S_{eff}$ is the effective spin and $T_0$ is the effective temperature. These parameters permit a phenomenological description of the antiferromagnetic Mn-Mn exchange interaction. For noninteracting paramagnetic Mn$^{2+}$ ions $S_{eff}=2.5$ and $T_0=0$~K. We use these values to fit the data for Zn$_{0.99}$Mn$_{0.01}$Se. For their values at higher Mn concentrations in Zn$_{1-x}$Mn$_x$Se we refer, e.g., to Fig.~4 of Ref.~\onlinecite{Keller2001}.

The degree of circular polarization of the PL is defined as 
\begin{equation}
P_{c}(B)=\frac{I^{+}(B)-I^{-}(B)}{I^{+}(B)+I^{-}(B)}.
\label{eq:Pc}
\end{equation}
Here, $I^{+}$ and $I^{-}$ denote the photoluminescence intensities detected in $\sigma^+$ and $\sigma^-$ polarization, respectively. In our simplified consideration, it is determined by the thermal population of excitons or charge carriers on the Zeeman split sublevels. Then, it can be described by 
\begin{equation}
P_{c}(B)=\frac{\tau}{\tau+\tau_s} \tanh \left( \frac{\Delta E_Z(B,T_{Mn})}{2 k_B T_X} \right)  .
\label{eq:DCP}
\end{equation}
Here, $\tau$ and $\tau_s$ are the exciton lifetime and the spin relaxation time, respectively. $T_X$ is the temperature of the thermalized excitons, contributing to the radiative recombination. In DMS, typically $\tau_s \ll \tau$ and $\tau/(\tau+\tau_s)\approx 1$. However, the validity of this simplified consideration needs to be examined for the specific conditions of the experiments and samples.

\subsection{Direct optical transition}
\label{sec:Direct}

\begin{figure}[hbt]
	\begin{center}
\includegraphics[width=0.48\textwidth]{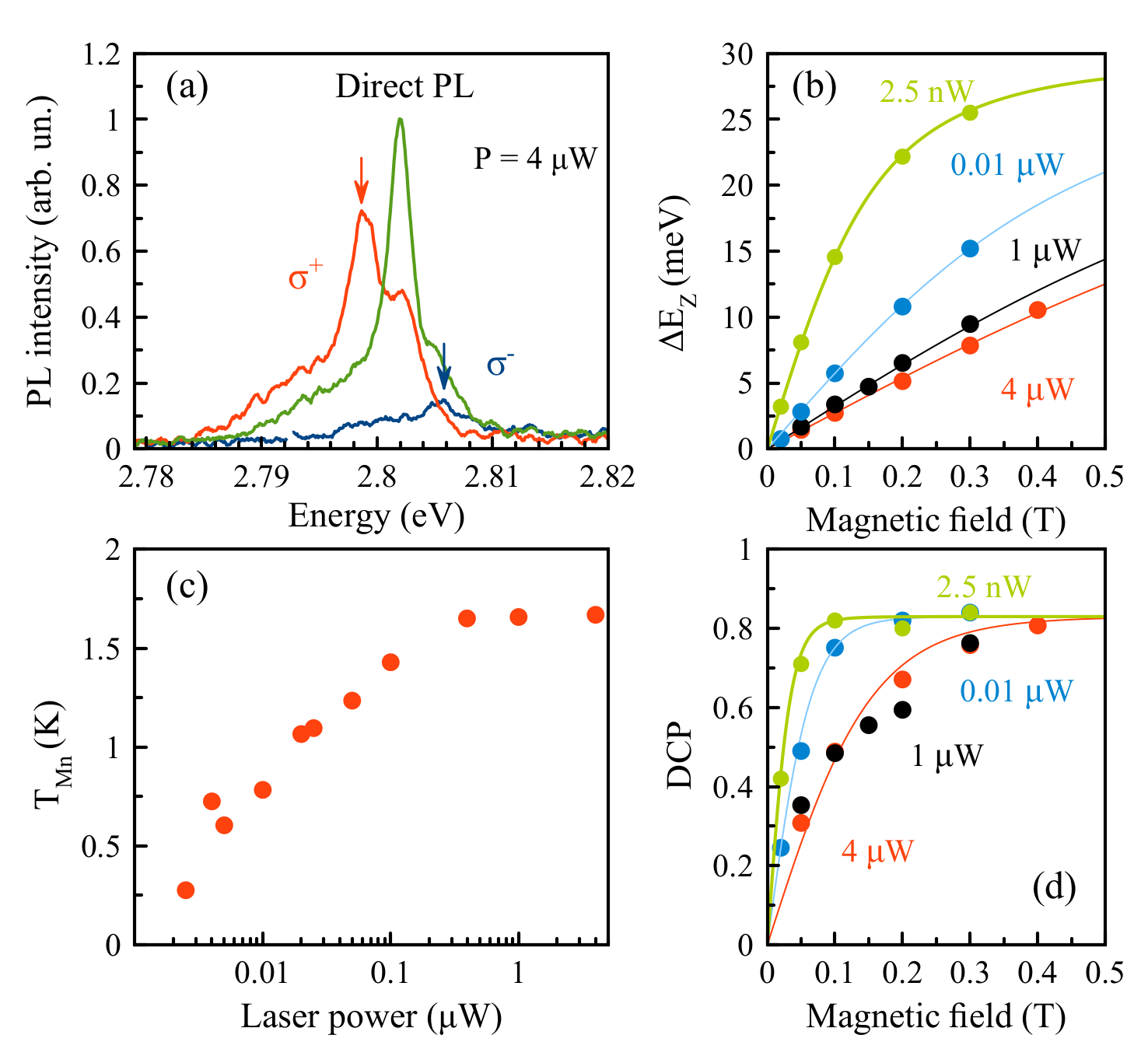}
\caption{Polarized photoluminescence of the Zn$_{0.99}$Mn$_{0.01}$Se/Be$_{0.93}$Mn$_{0.07}$Te QWs in magnetic field, measured at the direct PL energy at $T_s=16$~mK. 
(a) Direct PL band measured at  $B=0$~mT (green line) and $B=250$~mT in $\sigma^+$ (red) and $\sigma^-$ (blue) circular polarization. $P=4$~$\mu$W. Arrows mark exciton maxima split in magnetic field. 
(b) Magnetic field dependence of the giant Zeeman splitting  measured at various excitation powers (symbols). The lines are fits with Eq.~\eqref{eq:GZS1} using $S_{eff}=2.5$ and $T_0=0$~K. The parameter $T_{Mn}$ is evaluated from the fits.   
(c) Evaluated $T_{Mn}$ as function of the laser power for the base temperature $T_{s}=16$~mK.
(d) Magnetic field dependence of the circular polarization degree measured at various excitation powers (symbols). The lines are fits with Eq.~\eqref{eq:DCP} using the experimental values of $\Delta E_Z$ from panel (b). The fit parameters $T_X$ are 33~K, 22~K, and 24~K for the laser powers of 2.5~nW, 0.01~$\mu$W and 4~$\mu$W, respectively.
}
\label{Fig_1n}
\end{center}
\end{figure}

First, we present the results for the direct transition PL, where the magneto-optical properties are controlled solely by the Zn$_{0.99}$Mn$_{0.01}$Se layers. The direct PL measured at zero magnetic field for the base temperature of $T_s=16$~mK is composed of a narrow exciton line at 2.802~eV with full width at half maximum of 2.4~meV, see the green line in Fig.~\ref{Fig_1n}(a). In a magnetic field of $B=250$~mT, this line demonstrates a giant Zeeman splitting (GZS) of about 6~meV. Namely, the $\sigma^+$ polarized component (red spectrum) is shifted to lower energy and the $\sigma^-$ polarized component to higher energy. The magnetic field dependence of the GZS for the excitation power of $P=4$~$\mu$W is shown by the red symbols in Fig.~\ref{Fig_1n}(b). 

In order to decrease or even avoid laser heating of the lattice and the Mn spin system, we reduce the excitation power to 2.5~nW. Our limitation for the lowest possible applicable power comes from the duration of the signal accumulation time, which for $P=2.5$~nW is about one hour. To avoid accumulation of cosmic-ray spikes on the CCD detector, we perform 80 measurements with 50 s accumulation time each and sum up the results after elimination of these spikes. The GZS measured at various laser powers are plotted in Fig.~\ref{Fig_1n}(b). The lines are fits using Eq.~\eqref{eq:GZS1} with $N_0 \alpha$ and $N_0 \beta$ for Zn$_{1-x}$Mn$_x$Se as well as $S_{eff}=2.5$ and $T_0=0$~K. $T_{Mn}$ is evaluated as a fit parameter. Its dependence on excitation power is plotted in Fig.~\ref{Fig_1n}(c). One can see, that with decreasing power the temperature drops from 1.6~K down to 270~mK. Note, that even at the lowest power the Mn spin temperature exceeds significantly the base temperature achievable by the cryostat system of $T_s=16$~mK. Therefore, optical heating is still an important issue, which can be explained by the very long spin-lattice relaxation of the Mn$^{2+}$ ions in DMS with low Mn concentrations.

The circular polarization degree dependencies on magnetic field measured for the direct PL at various excitation powers are shown in Fig.~\ref{Fig_1n}(d). They demonstrate qualitatively the same trend as the GZS from Fig.~\ref{Fig_1n}(b), namely the DCP increases with decreasing power. With increasing magnetic field, the DCP grows linearly before reaching saturation at $P_c=0.84$. One would expect that 100\% of polarization can be reached, as typically observed in bulk Zn$_{1-x}$Mn$_x$Se. We suggest that the smaller saturation DCP is caused by some depolarization factors in the detection scheme. The lines in  Fig.~\ref{Fig_1n}(d) are fits made with Eq.~\eqref{eq:DCP} using the experimental values of $\Delta E_Z$ from panel (b). The obtained values of the fit parameter $T_X$ fall in the range of $22-33$~K, which are too high to be physically reasonable. We suggest that this indicates that in this case the DCP is determined by Mn spin fluctuations and magnetic polaron formation~\cite{DMS_book_Chap_7}, which considerably weaken down the DCP dependence on magnetic field. A detailed analysis of this mechanism is beyond the scope of this study.

\begin{figure}[hbt]
	\begin{center}
\includegraphics[width=0.48\textwidth]{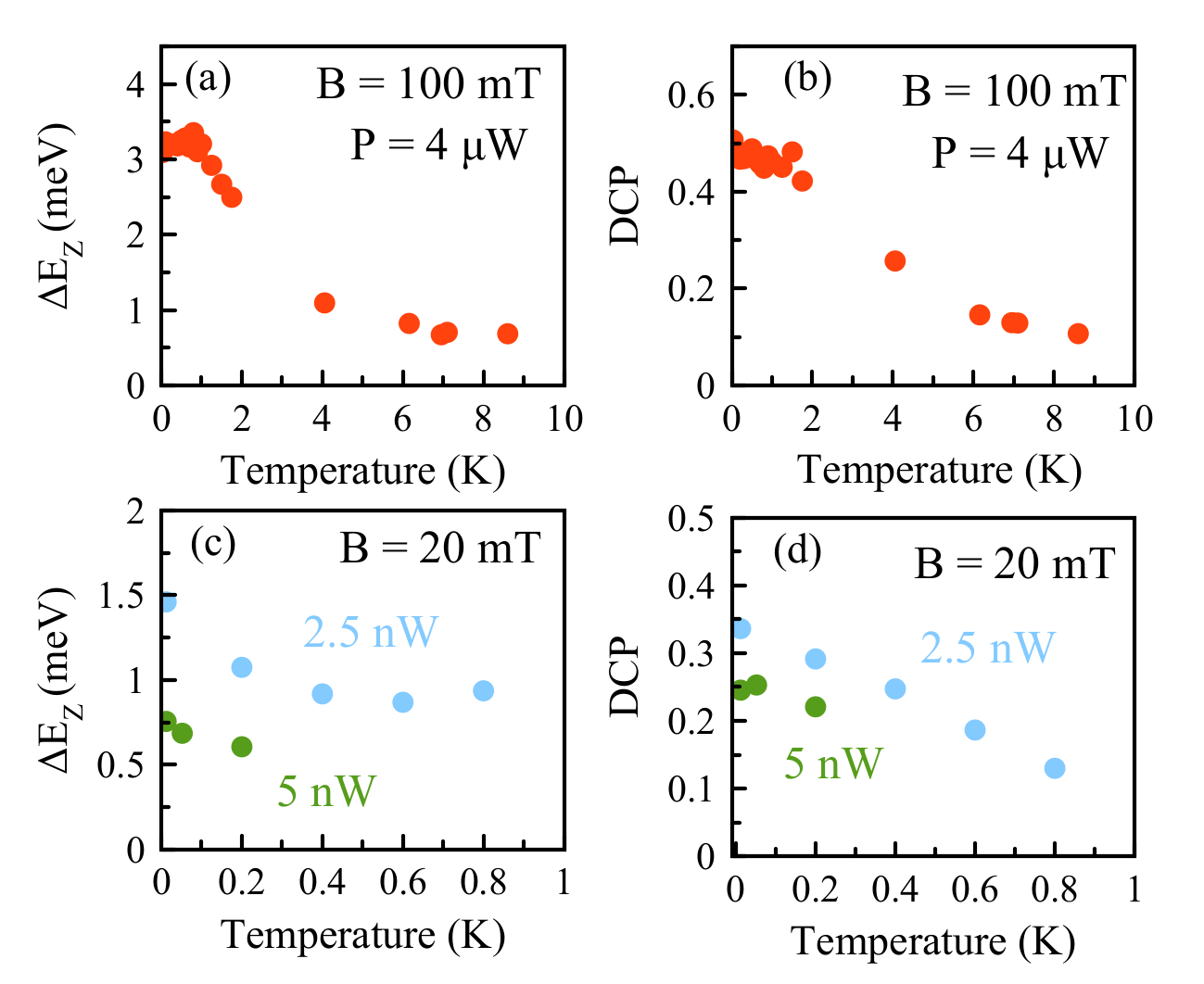}
\caption{Temperature dependences of the giant Zeeman splitting and the circular polarization degree of the direct PL. (a,b) Measurements at $B=100$~mT using $P=4$~$\mu$W. (c,d) Measurements at $B=20$~mT using $P=2.5$~nW (blue symbols) and 5~nW (green symbols). 
  	}
\label{Fig_2n}
	\end{center}
\end{figure}

Figure~\ref{Fig_2n} presents the results on the temperature dependence of the GZS and DCP. For this experiment, the base temperature in the cryostat is changed from 16~mK to 8.5~K, and monitored by a sensor located next to the sample. In Fig.~\ref{Fig_2n}(a,b), a relatively high excitation power of 4~$\mu$W is chosen, under such excitation the Mn spin temperature is 1.6~K even at the lowest base temperature $T_s=16$~mK. This explains why both the GZS and DCP remain about constant in the temperature range from 16~mK to about 1~K. A further temperature increase results in their reduction. For the smallest powers of 2.5 and 5~nW the temperature dependences of the GZS and DCP are given in the range from 16~mK to 1~K in Figs.~\ref{Fig_2n}(c,d). They are sensitive to the temperature increase starting from even the smallest elevation. For example, the DCP at $P=2.5$~nW decreases monotonically, evidencing that the Mn spin temperature increases steadily with $T_s$.

\subsection{Indirect optical transition}
\label{sec:Indirect}

\begin{figure}[hbt]
	\begin{center}
\includegraphics[width=0.45\textwidth]{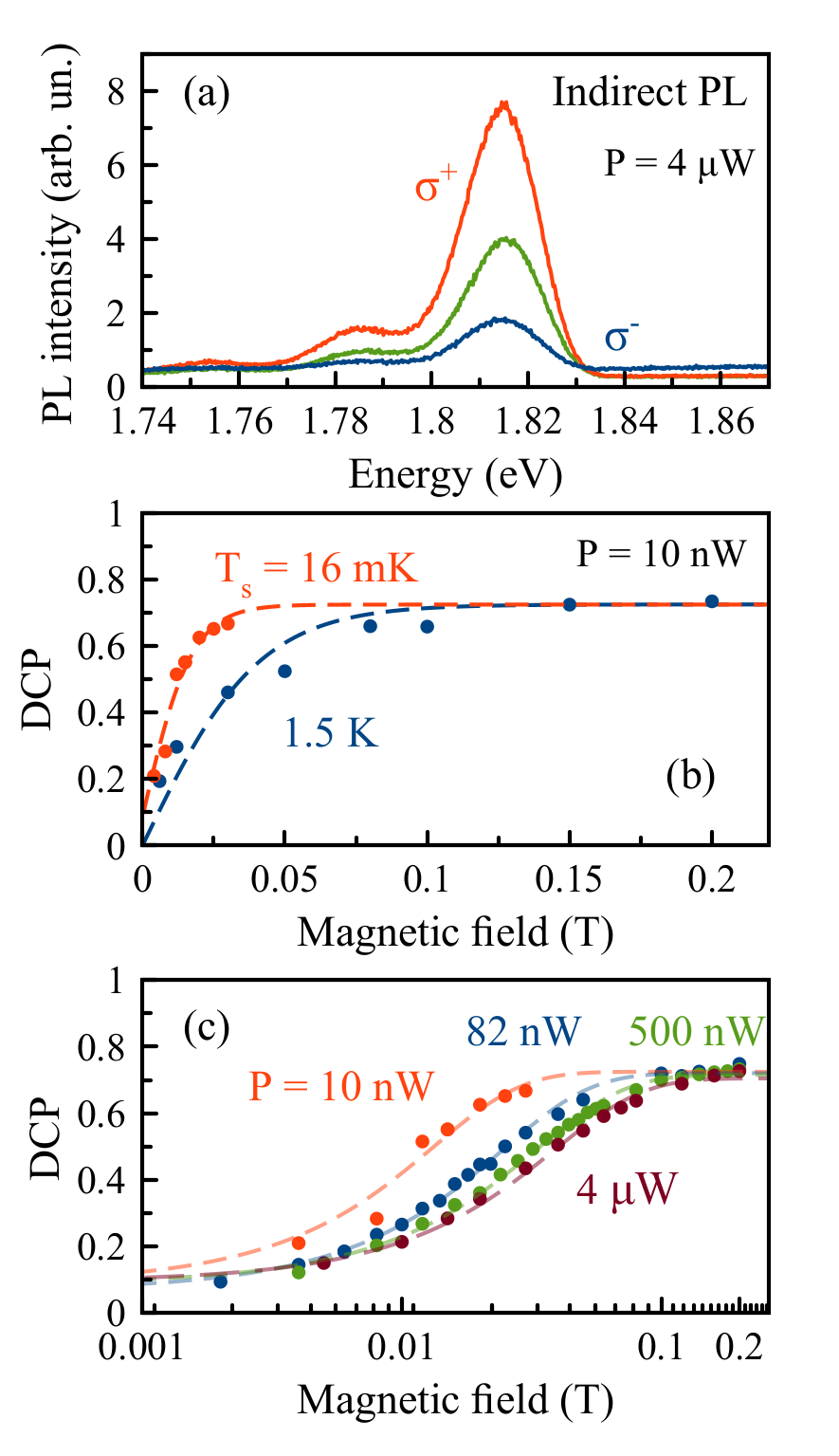}
\caption{Polarized photoluminescence of the Zn$_{0.99}$Mn$_{0.01}$Se/Be$_{0.93}$Mn$_{0.07}$Te QWs in magnetic field measured at the indirect PL band for $T_s=16$~mK.  
(a) Indirect PL band measured at  $B=0$~mT (green line) and $B=120$~mT in $\sigma^+$ (red) and  $\sigma^-$ (blue) circular polarization. $P=4$~$\mu$W.
(b)  Magnetic field dependence of the circular polarization degree measured at $T_s=16$~mK  and 1.5~K for $P=10$~nW. 
(c) Magnetic field dependences of the circular polarization degree measured at $T_s=16$~mK using various laser powers (symbols). Lines in panels (b,c) are guide to the eye.
}
\label{Fig_3n}
	\end{center}
\end{figure}

\begin{figure}[hbt]
	\begin{center}
\includegraphics[width=0.45\textwidth]{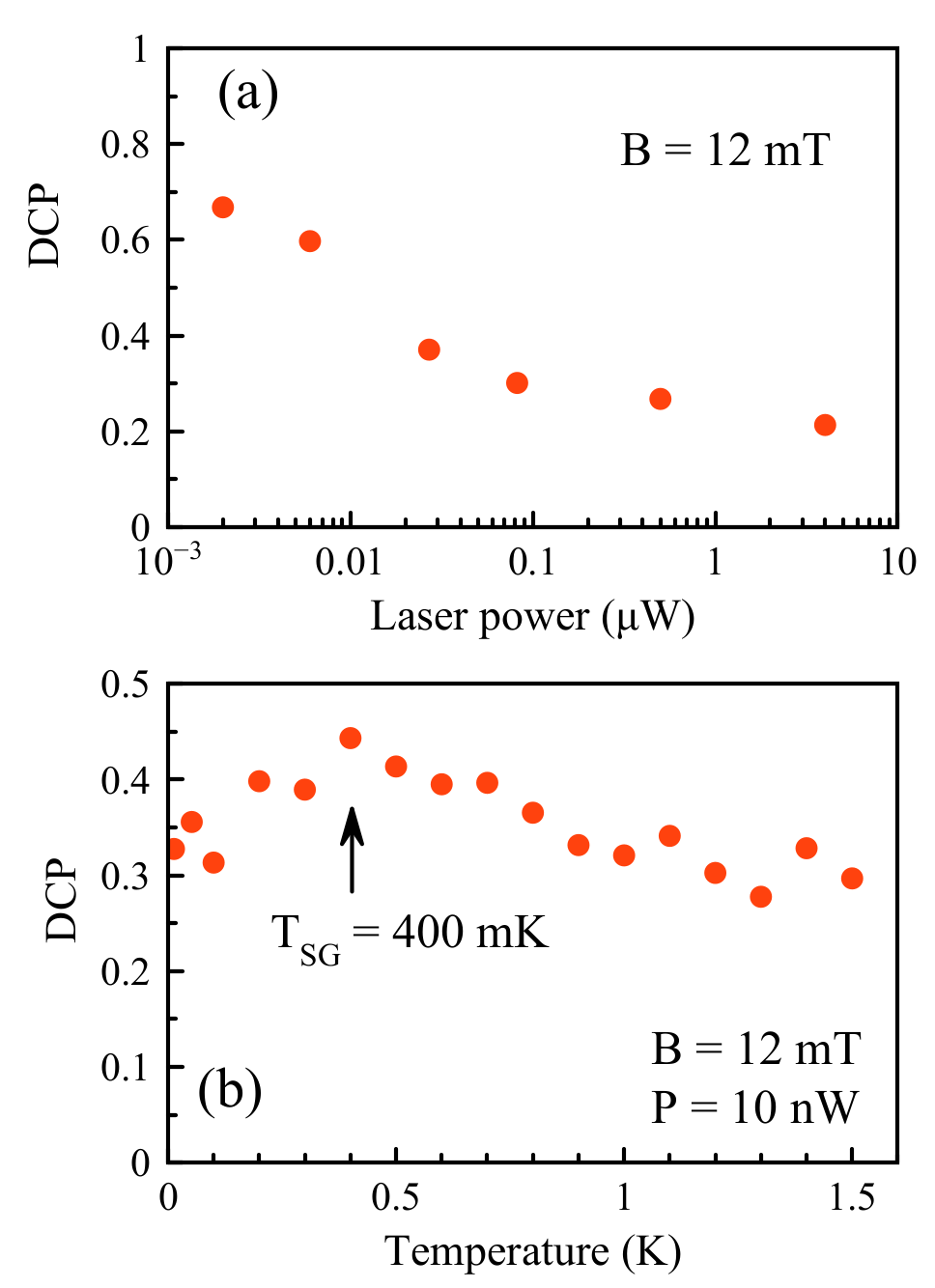}
\caption{Circular polarization degree of the indirect PL in a magnetic field of $B=12$~mT. 
(a) Power dependence of the DCP measured at $T_s=16$~mK. 
(b) Temperature dependence of the DCP measured at $P=10$~nW.
}
\label{Fig_4n}
	\end{center}
\end{figure}

The indirect recombination of the electrons in the Zn$_{0.99}$Mn$_{0.01}$Se layers with holes in the Be$_{0.93}$Mn$_{0.07}$Te layers provides the PL band at 1.815~eV (Fig.~\ref{Fig_structure} and the green line in Fig.~\ref{Fig_3n}(a)). In a magnetic field of 120~mT, it becomes circularly polarized, see the red and blue spectra for $\sigma^+$ and $\sigma^-$ polarization, respectively. 
The indirect PL band is spectrally broad so that its GZS cannot be well resolved at the low magnetic fields used in our studies. Therefore, in  Fig.~\ref{Fig_3n}(b,c) we show the results for the DCP and its magnetic field dependencies at various base temperatures and laser powers. As the peak PL intensity of the indirect PL is smaller than that of the direct one, the lowest laser power we can apply in these experiments is 10~nW. In Figure~\ref{Fig_3n}(b) two dependences for this power are shown for $T_s=16$~mK and 1.5~K. The DCP readily increases with sample cooling, while showing saturation at $P_c=0.7$. Figure~\ref{Fig_3n}(c) illustrates the impact of laser heating on the Mn spin system, by presenting the magnetic field dependencies of the DCP at various excitation powers, measured at $T_s=16$~mK. For better visualization, we show in  Fig.~\ref{Fig_4n}(a) the power dependence of the DCP measured at $B=12$~mT. 

Note that the polarization properties of the indirect emission in ZnSe/BeTe structures show a strongly specific feature: Here, an electron confined in the ZnSe layer recombines with a hole from the BeTe layer. As mentioned, the band offsets are very large reaching 2.2~eV in the conduction band and 0.8~eV in the valence band~\cite{Platonov1998}. As a result, the electron and hole wavefunctions overlap only within a short distance in the vicinity of the interface so that the chemical bonds and their orientation at the interface play an important role in the polarization properties, especially for this heterosystem, which has no common ion in ZnSe and BeTe layers. 

The related physics was examined experimentally and theoretically in Refs.~\cite{Platonov1999,Yakovlev2000,Platonov2002} for ZnSe/BeTe QWs and with the help of incorporating Mn$^{2+}$ ions, which allows us to reach a complete spin orientation of the electrons and holes~\cite{Yakovlev2002}. It was shown that the indirect emission at single interface has a linear polarization degree of about $P_l=0.70$ (i.e. 70\%), when the recombining carriers have no spin polarization (e.g. at zero magnetic field). Even for fully polarized electrons and/or holes, which in direct band gap DMS materials results in $P_c=1$, the circular polarization degree is limited to $P_c=0.70$ for the indirect emission. This is the result of the equation for the total polarization degree $P^2=P_c^2+P_l^2$~\cite{Platonov2002}. One can see in Fig.~\ref{Fig_3n}(b,c) that the DCP in the studied sample saturates at $P_c \approx 0.70$. 

The temperature dependence of the DCP of the indirect PL measured at the lowest applicable power of $10$~nW at $B=12$~mT is shown in Fig.~\ref{Fig_4n}(b). Surprisingly, this dependence differs considerably from the one measured at 2.5~nW for the direct PL, compare with the blue symbols in Fig.~\ref{Fig_2n}(d). Namely, the observed changes are weaker and the dependence has a maximum at about 400~mK, below which the DCP decreases with decreasing temperature. We attribute these features to hole spin polarization in the Be$_{0.93}$Mn$_{0.07}$Te layers. Note that the spin-lattice relaxation in DMS with a Mn content of $x=0.07$ is greatly faster than in a DMS with $x=0.01$. Therefore, a much weaker effect of laser heating is expected for the Mn system in the  Be$_{0.93}$Mn$_{0.07}$Te layers compared to the Zn$_{0.99}$Mn$_{0.01}$Se layers.  

We consider this behavior as evidence of spin glass formation~\cite{Yakovlev1999sg} in the Mn spin system of the  Be$_{0.93}$Mn$_{0.07}$Te layers with a critical temperature of $T_{SG}=400$~mK for this phase transition. This value is in reasonable agreement with the spin glass transitions detected via SQUID measurements of the magnetic susceptibility for DMS compounds with $x=0.07$: in  Cd$_{1-x}$Mn$_x$Te the critical temperature $T_{SG} = 230$~mK~\cite{Novak1985} and in Hg$_{1-x}$Mn$_x$Te, $T_{SG} = 600$~mK~\cite{Brandt1983}.  Note that spin glass formation has not been reported so far for Be$_{1-x}$Mn$_x$Te.

\section{Conclusions}

We have examined the magneto-optical properties of a Zn$_{0.99}$Mn$_{0.01}$Se/Be$_{0.93}$Mn$_{0.07} $Te DMS heterostructure with type-II band alignment at cryogenic temperatures down to 16~mK. For the sample fixed on a cold finger, the laser heating of the Mn spin system becomes a decisive factor. We have used the giant Zeeman splitting effect as a tool to evaluate the Mn spin temperature, reaching a minimum value of 270~mK at the lowest applicable excitation power of 2.5~nW. Evidence of spin glass formation in the Mn spin system of the  Be$_{0.93}$Mn$_{0.07}$Te layers with a critical temperature $T_{SG}=400$~mK has been found in the temperature dependence of the circular polarization degree of the indirect PL band.

\section*{ACKNOWLEDGMENTS}
This study was made possible through funding of the He3-He4 dilution refrigerator through a Major Research Instrumentation Proposal of the Deutsche Forschungsgemeinschaft (project number 427377618).

\section*{ORCID}

Dennis Kudlacik    0000-0001-5473-8383   \\  
Nataliia E. Kopteva  0000-0003-0865-0393 \\  
Mladen Kotur         0000-0002-2569-5051 \\  
Dmitri R. Yakovlev   0000-0001-7349-2745 \\  
Manfred Bayer        0000-0002-0893-5949 \\  

\end{document}